\documentclass[A4]{PoS}
\def \de {\partial}
\def \a {\alpha}
\def \b {\beta}
\def \g {\gamma}

\def \d {\delta}

\def \m {\mu}
\def \n {\nu}

\def \zh {\hat z}
\def \be {\begin{equation}}
\def \ee {\end{equation}}
\def \bea {\begin{eqnarray}}
\def \eea {\end{eqnarray}}
\def \non {\nonumber}
\def \noi {\noindent}
\def \ra {\rightarrow}

\def\laq{~\raise 0.4ex\hbox{$<$}\kern -0.8em\lower 0.62ex\hbox{$\sim$}~}
\def\gaq{~\raise 0.4ex\hbox{$>$}\kern -0.7em\lower 0.62ex\hbox{$\sim$}~}

\def \wt {\widetilde}

\title{Soft-wall model of AdS/QCD:\\ The case of light scalar mesons }

\ShortTitle{Soft-wall model of AdS/QCD: The case of light scalar
mesons }

\author{Fulvia De Fazio\\
        Istituto Nazionale di Fisica Nucleare, Sezione di Bari, Italy\\
        E-mail: \email{fulvia.defazio@ba.infn.it}}

\abstract{We study light scalar mesons   in the AdS/QCD soft-wall
model with a background dilaton field, to investigate the features
of this approach in describing QCD properties in the strong
coupling regime. We find that the masses and decay constants are
compatible with experiment and QCD determinations if $a_0(980)$
and $f_0(980)$ are identified as the lightest scalar mesons;
moreover, the states are organized in linear Regge trajectories
with  the same slope of vector mesons. Strong couplings of scalar
states to pairs of light pseudoscalar mesons turn out to be small,
at odds with experiment and QCD estimates:  this discrepancy is
related to the description of chiral symmetry breaking  in this
holographic model.}

\FullConference{8th Conference Quark Confinement and the Hadron Spectrum\\
         September 1-6, 2008\\
         Mainz. Germany}

\begin{document}

The  AdS/CFT correspondence  was  proposed by Maldacena
\cite{Maldacena:1997re} as a duality between a type IIB string
theory defined on AdS$_5$ $\times S^5$ space (AdS$_5$ being a 5
dimensional Anti de Sitter space and $S^5$ the 5d sphere) and a
${\cal N}$=4 super Yang-Mills theory with gauge group $SU(N_c)$,
for large $N_c$. Later on,  it was supposed that the
correspondence could be generalized as an equivalence between a
theory defined on AdS$_{d+1}\times{\cal C}$ (${\cal C}$ being a
compact manifold) and a conformal field theory living on the flat
boundary ${\cal M}_d$ of the AdS space \cite{Witten:1998qj}. This
has provided new hints on the possibility of describing strong
interaction processes by string-inspired approaches. Two main ways
are followed to achieve such a result. The first one, the
so-called top-down approach, consists in starting from a string
theory trying to derive a low-energy QCD-like theory on  ${\cal
M}_d$ through compactifications of the extra dimensions
\cite{top-down}. In the second one, the  bottom-up approach, one
starts from $4d$ QCD and attempts to construct its higher
dimensional dual theory \cite{polchinsky}, with phenomenological
properties as guidelines.

In both approaches it is necessary  to break conformal invariance,
since QCD is not a conformal theory \cite{confqcd}, and to account
for confinement. In the bottom-up approach, one possibility
(hard-wall model) is to use a five dimensional ``AdS-slice''
 with the fifth (holographic) coordinate
$z$ varying  up to $z_{max}$ of ${\cal
O}(\frac{1}{\Lambda_{QCD}})$ \cite{polchinsky,son1}.
 Another proposal to break conformal invariance consists in
introducing in the $5d$ AdS  space a background dilaton field
(soft-wall model) \cite{Andreev soft wall,son2}.

Here we report an analysis  of  light scalar mesons  in the
 soft wall model \cite{Colangelo:2008us}.  In
particular, we describe the derivation of the mass spectrum, the
decay constants and the strong couplings of scalar mesons to pairs
of light pseudoscalars. The comparison of the results obtained in
the AdS framework with experiment and QCD calculations can shed
light on the features and drawbacks of the model. Other analyses
of scalar mesons in  holographic approaches can be found in
\cite{pomarol2,vega}.

\section{The model}\label{model}

In the $5d$ space we consider the metric: $ ds^2=g_{MN} dx^M
dx^N=\frac{R^2}{z^2}\,\big(\eta_{\m\n}dx^{\m}dx^{\n}+dz^2\big) $
with $\eta_{\m\n}=\mbox{diag}(-1,+1,+1,+1)$; $R$ is the AdS
curvature radius and  the holographic coordinate $z$ runs from
zero to infinity. The model is characterized by a background
dilaton field: $ \Phi(z)=(c z)^2$, the form of which is chosen  to
obtain light vector mesons with linear Regge trajectories
\cite{son2}; $c$ is a dimensionful parameter setting the scale of
QCD quantities. We consider the $5d$ action:
\begin{equation}
S_{eff}=-\frac{1}{k}\int
d^5x\sqrt{-g}\,e^{-\Phi(z)}\,\mbox{Tr}\Big\{|DX|^2+m_5^2X^2+\frac{1}{2g_5^2}\big(F_V^2+F_A^2\big)\Big\}
\label{action1}
\end{equation}
where $g$ is the determinant of the metric tensor $g_{MN}$
 and $k$  a dimensionful parameter
providing a dimensionless action. This action includes  fields
dual to QCD operators defined at the boundary $z=0$. There is a
scalar bulk field $X$, whose mass  is fixed by the AdS/CFT
relation: $m_5^2R^2=(\Delta-p)(\Delta+p-4)$,  $\Delta$ being the
dimension of the $p-$form QCD operator dual to $X$. This field,
written as $ X=(X_0+S)e^{2i\pi}$, contains a  field
$X_0(z)=\frac{v(z)}{2}$, the  scalar field $S(x,z)$ and the chiral
field $\pi(x,z)$.  $X_0$ is  dual to $\langle \bar q q\rangle$ and
is responsible for
   chiral symmetry breaking. $S$ includes
singlet $S_1(x,z)$ and  octet $S_8^a(x,z)$ components:
$S=S^AT^A=S_1T^0+S_8^aT^a $ with $T^0={1}/\sqrt{2n_F}$ and $T^a$
the generators of $SU(3)_F$ ($A=0,a$, and $a=1,\ldots 8$). $S^A$
is  dual  to  ${\cal O}^A_S(x)=\overline{q}(x)T^A q(x)$, so that
$\Delta=3$, $p=0$ and  $m_5^2R^2=-3$. The  fields $A^a_{L,R}(x,z)$
are introduced to gauge the chiral symmetry in the $5d$ space.
They are dual to  $\bar q_{L,R} \gamma_\mu T^a q_{L,R}$ and can be
written
 in
terms of vector $V$ and axial-vector  $A$ fields: $V^M=\frac{1}{2}
(A_L^M+A_R^M)$,  $A^M=\frac{1}{2} (A_L^M-A_R^M)$, so that $
F_{V}^{MN}=\partial^M V^N - \partial^N V^M-i[V^M,V^N]-i[A^M,A^N]$,
$ F_{A}^{MN}=\partial^M A^N -
\partial^N A^M-i[V^M,A^N]-i[A^M,V^N]$
and $D^MX=\partial^M X-i [V^M ,X] -i \{A^M,X\}$.

Using $S_{eff}$ in (\ref{action1}), we assume that the AdS/CFT
duality relation: $\displaystyle{
  \left\langle e^{i\int d^4x\,{\cal
  O}(x)f_0(x)}\right\rangle_{QCD}=e^{iS_{eff}}}$ holds,
where the \emph{lhs} is the QCD generating functional in which the
sources $f_0(x)$ of the $4d$ ${\cal O}(x)$ operators are the
boundary ($z\ra0$) limits of the corresponding (dual) $5d$ fields.
We then derive the properties of light scalar mesons  in  the
soft-wall model.

\section{Spectrum of scalar mesons}
Let us consider the quadratic part of the action (\ref{action1})
involving the scalar fields $S^A(x,z)$:
\begin{equation}
S^{(2)}_{eff}=-\frac{1}{2k}\int
d^5x\sqrt{-g}\,e^{-\Phi(z)}\,\Big(g^{MN}\de_MS^A\de_NS^A+m_5^2S^AS^A\Big)
\,\,\, .
\end{equation}
From this term, we derive  the equation of motion  for the field
$S^A$:
\begin{equation}
\eta^{MN}\de_M\Big(\frac{R^3}{z^3}\,e^{-\Phi(z)}\de_NS\Big)+3\frac{R^3}{z^5}\,e^{-\Phi(z)}S=0
\, \label{eq-mot}
\end{equation}
where we dropped the flavour index $A$. We identify scalar meson
states with the normalizable solutions of this equation
corresponding to the discrete mass spectrum \cite{vega}: $
-q^2_n=m_n^2=c^2(4n+6) $ with  integer $n$, and eigenfunctions
expressed in terms of the generalized Laguerre polynomials
$\displaystyle{ \tilde S_n(\zh)=\sqrt{\frac{2}{n+1}}\,\zh^3
L_n^1(\zh^2)}$ (we have defined $S(x,z)=\int \frac {d^4 q}{(2
\pi)^4}\,e^{i q \cdot x} \tilde S(q,z)$).

Scalar mesons are organized in linear Regge trajectories with the
same slope  as  vector mesons and scalar glueballs, the spectral
condition of which is $ m_{\rho_n}^2=c^2(4n+4)$ \cite{son2} and
$m_{G_n}^2=c^2(4n+8)$ \cite{jugeau1}, respectively, with  the
parameter $c$ setting the scale of all hadron masses.

Scalar mesons turn out to be  heavier than vector mesons. This is
in quantitative agreement with experiment if $a_0(980)$ and
$f_0(980)$ are identified as the lightest scalar mesons, since the
results in AdS
$R_{f_0(a_0)}=\frac{m_{f_0(a_0)}^2}{m_{\rho^0}^2}=\frac{3}{2}$
agree with  $R_{f_0}^{exp}=1.597 \pm 0.033$ and
$R_{a_0}^{exp}=1.612 \pm 0.004$. Finally, scalar mesons are
lighter than scalar glueballs:
$\frac{m_{G}^2}{m_{f_0}^2}=\frac{4}{3}$ for the lowest-lying
states.

\section{Two-point  correlation function of the scalar operator}
Let us consider in QCD the two-point correlation function:
\begin{equation}
\Pi_{QCD}^{AB}(q^2)= i \int d^4x\,e^{i q \cdot x} \langle
0|T[{\cal O}^A_S(x) {\cal O}^B_S(0)]|0\rangle \label{twopt}
\end{equation}
with ${\cal O}^A_S(x)=\overline{q}(x)T^Aq(x)$. The AdS/CFT method
relates  this correlation function to the two-point correlator
obtained from the action (\ref{action1}): $
\displaystyle{\Pi_{AdS}^{AB}(q^2)=\d^{AB}\frac{R^3
c^4}{k}\,S(\frac{q^2}{c^2},\zh^2)\frac{e^{-\Phi(\zh)}}{\zh^3}\,
\de_{\hat z} S(\frac{q^2}{c^2},\zh^2)\Big|_{\zh\to0} }$ where the
bulk-to-boundary propagator $\wt S(q^2,z)=S(q^2/c^2,\zh^2)\wt
S_0(q^2)$ is obtained solving eq. (\ref{eq-mot}) for all
four-momenta $q^2$. We get:
\begin{eqnarray}
\Pi^{AB}_{AdS}(q^2)&=&\d^{AB}\frac{4c^2R}{k}\,\biggl[\left(\frac{q^2}{4c^2}+\frac{1}{2}\right)\ln(c^2z^2)+
\left(\g_E-\frac{1}{2}\right)+\frac{q^2}{4 c^2}\left(2\g_E-\frac{1}{2}\right)\non \\
&+&\left(\frac{q^2}{4c^2}+\frac{1}{2}\right)\,\psi(\frac{q^2}{4c^2}+\frac{3}{2})\biggr]\biggr|_{\,z=z_{min}}
\,\,\, \label{twoptAdS1}
\end{eqnarray}
(omitting a ${\cal O}(\frac{1}{z^2})$ contact term). This
expression   shows the presence of a discrete set of poles, those
the Euler function $\psi$, with masses   $m_n^2=c^2(4n+6) $ and
residues $ F_n^2=\frac{R}{k}\,16\,c^4(n+1)$. The factor
$\frac{R}{k}$ can be fixed by matching  (\ref{twoptAdS1}) in the
$q^2\to +\infty$ (i.e. in the short-distance) limit, expanded in
powers of $1/q^2$,  with the QCD result, giving: $
\frac{R}{k}=\frac{N_c}{16\pi^2}$. The residues of the two-point
correlator, related to the scalar meson decay constants, are now
determined: $ F_n^2=\frac{N_c}{\pi^2}\,c^4(n+1)$.

We compare this result to QCD calculations. For $a_0(980)$, the
following result was obtained  $ F_{a_0}=\langle 0| {\cal
O}^3_S|\,a_0(980)^0\rangle = (0.21 \pm 0.05) $  GeV$^2$
\cite{gokalp}.
 The AdS prediction is:
$F_{a_0}=\frac{\sqrt 3}{\pi}c^2=0.08$~GeV$^2$, having fixed $c$
from the $\rho^0$ mass: $c=\frac{m_\rho}{2}$. For the $f_0(980)$ a
similar result was obtained for the matrix element of the $s\bar
s$ operator: $\langle 0|\,\bar s s\,| f_0(980)\rangle = (0.18 \pm
0.015)$~GeV$^2$ \cite{defazio}.  Considering the uncertainties in
QCD determinations, the  AdS results differ by about a factor of
two.

\section{Interaction of scalar mesons  with a pair of pseudoscalar mesons}\label{sec:inter}

In   (\ref{action1}) the interaction terms involving one scalar
$S$ and two light pseudoscalar fields $P$  only appear in the term
Tr$\left\{|DX|^2\right\}$. Using  the equations of motion and
writing the axial-vector bulk field in terms of the transverse and
longitudinal components: $A_M=A_{\perp\,M}+\de_M\phi$, we have:
\begin{equation}\label{spipi}
S^{(SPP)}_{eff}=-\frac{4}{k}\int d^5x\sqrt{-g}\,e^{-\Phi(z)}g^{MN}
v(z)\,\mbox{Tr}\Big\{S(\de_M\pi-\de_M\phi)(\de_N\pi-\de_N\phi)\Big\}\,\,.
\end{equation}
Defining $\psi^a=\phi^a-\pi^a$ and for $n_F=3$  we have:
\be S^{(SPP)}_{eff}=-\frac{R^3}{k}\int
d^5x\,\frac{1}{z^3}\,e^{-\Phi(z)}v(z)\left[
\frac{2}{\sqrt{6}}\,S_1\,\eta^{MN}(\de_M\psi^a)(\de_N\psi^a)
+d^{abc}\,S_8^a\,\eta^{MN}(\de_M\psi^b)(\de_N\psi^c) \right] \,\,
. \ee
If the fields are expressed in terms of their bulk-to-boundary
propagators and of the corresponding sources,  the three point
function involving two pseudoscalar and one scalar operator can be
obtained by functional derivation of the action with respect to
the source fields. The AdS result can be compared with the QCD
three-point
 function:
\begin{equation}\label{3ptQCD}
\Pi_{QCD\a \b}^{abc}(p_1,p_2)=i^2 \int d^4x_1 d^4x_2\,e^{i
p_1\cdot x_1} e^{i p_2\cdot x_2} \langle0|T[{\cal
O}^b_{5_{\a}}(x_1){\cal O}^a_S(0) {\cal
O}^c_{5_{\b}}(x_2)]|0\rangle \,\,\,.
\end{equation}
Since $\Pi_{QCD\a \b}^{abc}(p_1,p_2)$ can be written in terms of
the coupling $g_{S_nPP}$ as follows:
\begin{equation}
\Pi^{abc}_{QCD \a \b}(p_1,p_2)=
d^{\,abc}\frac{p_{1\a}p_{2\b}}{p_1^2p_2^2}f_{\pi}^2\,\sum_{n=0}^{\infty}\frac{F_ng_{S_nPP}}{q^2+m_n^2}
\,\,\,
\end{equation}
with $q=-(p_1+p_2)$ and $f_{\pi}$ the pion decay constant,
comparison with the  expression obtained in AdS allows to
determine $g_{S_nPP}$. For the lowest radial number $n=0$ we have:
\begin{equation}\label{coupl}
g_{S_0PP}=\frac{\sqrt{N_c}}{4\pi}\frac{m_{S_0}^2}{f_{\pi}^2} Rc
\int_{0}^{\infty}d\hat{z}\,e^{-\hat{z}^2} v(\hat{z})  \,\,\,
\end{equation}
with $\hat{z}=cz$. $g_{S_0PP}$ depends linearly on the  field
$v(z)$ introduced in Section \ref{model}, which can be obtained
solving the equation of motion which stems from the action
(\ref{action1}).  The numerical result is small, of ${\cal O}(10)$
MeV depending on the input quark mass, while phenomenological
determinations of  $g_{S_0PP}$ indicate sizeable values. For
example the experimental value of $g_{a_{0}\eta \pi}$ is:
$g_{a_{0}\eta \pi}=12\pm6$ GeV, while for $f_0$  a QCD estimate
gives: $g_{f_0 K^+ K^-}\simeq 6-8$ GeV \cite{defazio3}. The origin
of the small value for the $SPP$ couplings in the  soft-wall model
is related to the difficulty of correctly reproducing chiral
symmetry breaking in this model through the non vanishing chiral
condensate and light quark masses \cite{son2,Colangelo:2008us}.

\section{Conclusions}
In the  AdS soft-wall model   the masses of  scalar meson
 are close to experiment, but their decay
constants differ from the available  QCD determinations by about a
factor of two. The strong couplings $g_{SPP}$ are smaller than in
phenomenological determinations; this is related to a  difficulty
in precisely describing  chiral symmetry breaking within this
model.
\par
\noi {\bf Acknowledgments}\\
\noi  I thank P. Colangelo, F. Giannuzzi, F. Jugeau and S. Nicotri
for collaboration. This work was supported in part by the EU
Contract No. MRTN-CT-2006-035482, "FLAVIAnet".


\begin{thebibliography}{99}
\bibitem{Maldacena:1997re}
  J.~M.~Maldacena,
  {\it The large N limit of superconformal field theories and supergravity},
  Adv.\ Theor.\ Math.\ Phys.\  {\bf 2}  (1998) 231
  [Int.\ J.\ Theor.\ Phys.\  {\bf 38} (1999) 1113].
%
\bibitem{Witten:1998qj}
  E.~Witten,
 {\it Anti-de Sitter space and holography},
  Adv.\ Theor.\ Math.\ Phys.\  {\bf 2}  (1998) 253;
  {\it Anti-de Sitter space, thermal phase transition, and confinement in  gauge
  theories},
  ibidem  {\bf 2}  (1998) 505;
  S.~S.~Gubser {\it et al.},
  {\it Gauge theory correlators from non-critical string theory},
  Phys.\ Lett.\  B {\bf 428}  (1998) 105.


\bibitem{top-down}
For a recent review see:
  J.~Erdmenger, N.~Evans, I.~Kirsch and E.~Threlfall,
  {\it Mesons in Gauge/Gravity Duals - A Review},
  Eur.\ Phys.\ J.\  A {\bf 35} (2008) 81
 and references therein.

\bibitem{polchinsky}
  J.~Polchinski {\it et al.},
  {\it Hard scattering and gauge/string duality},
  Phys.\ Rev.\ Lett.\  {\bf 88} (2002)  031601.

\bibitem{confqcd}
For a  discussion see
 S.~J.~Brodsky and G.~F.~de Teramond,
  arXiv:0802.0514.


\bibitem{son1}
    J.~Erlich, E.~Katz, D.~T.~Son and M.~A.~Stephanov,
 {\it QCD and a holographic model of hadrons},
  Phys.\ Rev.\ Lett.\  {\bf 95} (2005)  261602;
   L.~Da Rold and A.~Pomarol,
{\it Chiral symmetry breaking from five dimensional spaces},
  Nucl.\ Phys.\  B {\bf 721} (2005) 79.

\bibitem{Andreev soft wall}
  O.~Andreev,
{\it $1/q^2$ corrections and gauge / string duality},
  Phys.\ Rev.\  D {\bf 73} (2006) 107901.

\bibitem{son2}
A.~Karch {\it et al.},
 {\it Linear confinement and AdS/QCD},
  Phys.\ Rev.\  D {\bf 74}  (2006) 015005.

\bibitem{Colangelo:2008us}
P.~Colangelo, F.~De Fazio, F.~Giannuzzi, F.~Jugeau and S.~Nicotri,
{\it Light scalar mesons in the soft-wall model of AdS/QCD},
  Phys.\ Rev.\  D {\bf 78} (2008) 055009.

\bibitem{pomarol2}
  K.~Ghoroku, N.~Maru, M.~Tachibana and M.~Yahiro,
  {\it Holographic model for hadrons in deformed AdS(5) background},
  Phys.\ Lett.\  B {\bf 633}, 602 (2006);
    L.~Da Rold and A.~Pomarol,
  {\it The scalar and pseudoscalar sector in a five-dimensional approach to
  chiral symmetry breaking},
  JHEP {\bf 0601}(2006) 157;
    H.~Forkel, M.~Beyer and T.~Frederico,
  {\it Linear square-mass trajectories of radially and orbitally excited hadrons
  in holographic QCD},
  JHEP {\bf 0707} (2007) 077;
T.~Huang and F.~Zuo,
  {\it Couplings of the Rho Meson in a Holographic dual of QCD with Regge
  Trajectories},
  Eur.\ Phys.\ J.\  C {\bf 56} (2008) 75.


\bibitem{vega}
   A.~Vega and I.~Schmidt,
{\it Scalar hadrons in $AdS_{5} \times S^{5}$},
  Phys.\ Rev.\  D {\bf 78} (2008) 017703.

\bibitem{jugeau1}
  P.~Colangelo, F.~De Fazio, F.~Jugeau and S.~Nicotri,
  {\it On the light glueball spectrum in a holographic description of QCD},
  Phys.\ Lett.\  B {\bf 652} (2007) 73.


\bibitem{gokalp}
  A.~Gokalp, Y.~Sarac and O.~Yilmaz,
{\it Scalar $a_0$-meson contributions to radiative $\omega \to$
$\pi_0 \eta \gamma$ and
  $\rho_0 \to \pi_0 \eta \gamma$ decays},
  Eur.\ Phys.\ J.\  C {\bf 22} (2001) 327.

\bibitem{defazio}
  F.~De Fazio and M.~R.~Pennington,
 {\it Probing the structure of $f_0(980)$ through radiative $\Phi$ decays},
  Phys.\ Lett.\  B {\bf 521} (2001) 15.

\bibitem{defazio3}
  P.~Colangelo and F.~De Fazio,
{\it Coupling $g_{f_0 K^+ K^-}$ and the structure of $f_0(980)$},
  Phys.\ Lett.\  B {\bf 559} (2003) 49.

\end{thebibliography}
\end{document}